\documentclass[fleqn,10pt]{wlscirep}

\usepackage{subfig}
\usepackage{multirow}
\usepackage{rotating}
\usepackage[utf8]{inputenc}

\title{Investigating the role of musical genre in human perception of music stretching resistance
\thanks{If you are interested in the collected data, please contact Dr.~Chen: chenjun14@mails.thu.edu.cn.}}

\author[+,1]{Jun Chen}
\author[*,1]{Chaokun Wang}
\affil[1]{School of Software, Tsinghua University, Beijing, 100084, P.R. China.}
\affil[+]{chenjun14@mails.thu.edu.cn}
\affil[*]{chaokun@tsinghua.edu.cn}

\begin{abstract}
To stretch a music piece to a given length is a common demand in people's daily lives,
e.g., in audio-video synchronization and animation production.
However,
it is not always guaranteed that the stretched music piece is acceptable for general audience
since music stretching suffers from people's perceptual artefacts.
Over-stretching a music piece will make it uncomfortable for human psychoacoustic hearing.
The research on music stretching resistance attempts to estimate the maximum stretchability of music pieces to further avoid over-stretch.
It has been observed that musical genres can significantly improve the accuracy of automatic estimation of music stretching resistance,
but how musical genres are related to music stretching resistance has never been explained or studied in detail in the literature.
In this paper, the characteristics of music stretching resistance are compared across different musical genres.
It is found that music stretching resistance has strong intra-genre cohesiveness and inter-genre discrepancies in the experiments.
Moreover, the ambiguity and the symmetry of music stretching resistance are also observed in the experimental analysis.
These findings lead to a new measurement on the similarity between different musical genres based on their music stretching resistance.
In addition, the analysis of variance (ANOVA) also supports the findings in this paper by verifying the significance of musical genre in shaping music stretching resistance.
\end{abstract}
\begin{document}

\flushbottom
\maketitle
\thispagestyle{empty}

\section*{Introduction}

Music stretching resistance (\emph{abbr}. MSR)
which describes the acceptable range of time stretching rate of music piece for people's psychoacoustic hearing~\cite{wsola:icassp93,ola:sc00},
consists of the minimum compressing rate, denoted as $\alpha_{min}$ ($0.0<\alpha_{min}<1.0$),
and the maximum elongating rate, denoted as $\alpha_{max}$ ($1.0<\alpha_{max}<2.0$)~\cite{msr:spl13}.
The research into MSR is of broad interest in fields like time-scale modification of speech~\cite{wsola:icassp93,ola:sc00},
music resizing~\cite{musicresize:mms13,musicresize:mm12,musicresize:icme10}, 
dynamic music re-scaling~\cite{musicrescale:eurographics13} as well as other fields related to human perception of music and psychoacoustic hearing.
The computational method~\cite{msr:spl13} which estimates MSR by incorporating sound features (e.g.~spectral analysis, timbre, pitch) and musical genres,
has shown that musical genre is much more important in affecting MSR compared with sound features like timbre, pitch and rhythm.
However, there is still no in-depth research to investigate the relationship between MSR and musical genres to the best of our knowledge.

Generally speaking, the existence of MSR can be attributed to the process of receiving and recognizing accelerated and decelerated sounds where people would make positive or negative reaction to accept/reject the changes based on the satisfaction of their psychoacoustic hearing.
Basically, MSR is related to the perception of music and the artefacts of digital signal processing.
Human perception of music has much in common with human recognition of natural languages,
where the structures (syntax/harmony), the vocabularies (words/chords), the tonal properties (inflection/timbre) and the temporal features (prosody/rhythm) are shared~\cite{memorymanipulation:icmpc06}.
Meanwhile, the functional magnetic resonance imaging (fMRI) shows greater neuronal activity in the voice-selective regions of participants when they are listening to vocal sounds than to non-vocal sounds~\cite{voiceselective:nature00},
which indicates that people may also have different degrees of sensitivity towards music with/without lyrics, or with sparse/dense lyrics.
This difference may contribute to the creation of MSR since stretching operations change the density of lyrics (e.g.~words per minute), too.
The compressing or elongating operations~\cite{musicresize:mms13,musicresize:mm12,musicresize:icme10,wsola:icassp93,ola:sc00} on a given music piece change the tempos while preserving the pitch features~\cite{wsola:icassp93,ola:sc00,speedinmusic:jasa10},
which leads to a range of acceptability on user preferred speeds~\cite{speedmelody:sysmu08} as well as user tolerable speed ranges.
These are the evidences why MSR exists for general audience.

The dynamic attending theories~\cite{regularity:mathbehavior02,dynamicattending:psychreview89,rhythmicattending:cognition00,dynamicattending:psychreview99} posit that the internal mechanism, like a clock, a time keeper or an oscillator inside human beings, resonates and synchronizes with the periodicity of stretched music pieces to enable the audience to follow changes in tempo.
Without the knowledge of MSR, however, perceptual artefacts~\cite{musicresize:icme10} are more likely to occur
and degrade the auditory experience of general listeners
when a music piece is stretched at a rate out of its acceptable stretching range, i.e., overly compressed or overly elongated.
Most listeners are likely to identify the base tempo of music around $120$ BPM (beats per minute),
and the acceleration or the deceleration in speed usually induces more ambiguity in the identification of tempo~\cite{tempopercept:icmpc04,resonancetheory:cim04}.
To some extent, the na\"{i}ve approximation of acceptable stretching range, i.e., $\pm20\%$ (namely $\alpha_{min}=0.80$, $\alpha_{max}=1.20$)~\cite{msrbound:nime04}, is baseless and fails to utilize the features of music pieces,
considering that the perception of tempo appropriateness for a given music piece is determined by its contents~\cite{tempopercept:perception06}.
One of the categorical features of music is genre which incorporates cultural backgrounds and emotions of artists,
and characterizes the similarity between music pieces~\cite{genreclassify:spm06}.
Meanwhile, musical genres are also related to music content features, e.g., timbre, pitch and rhythm,
so that the automatic genre classification algorithms~\cite{genreclassify:spm06,genreclassify:sigir03,genreclassify:tsp02,genreclassify:spl07} could work,
which makes genre a potential factor in studying MSR.

In Chen's previous work~\cite{msr:spl13},
MSR values are discretized as labels along the axis of stretching rate.
The estimation of MSR is performed by classifying the label of the given music piece using its sound features (spectral analysis, timbre, pitch and tempo) and musical genre with the machine learning techniques.
It is observed that musical genre has larger contribution to improve the classification accuracy compared with sound features.
But how musical genre is related to MSR has never been studied in detail.
We believe that it is necessary to further explore the relationship between MSR and musical genres.

In this paper, we investigate the important role of musical genres in shaping MSR.
We find that MSR tends to be constant with small fluctuation within a given musical genre.
The significance of musical genres in the analysis of variance substantiates the existence of inter-genre discrepancy and intra-genre cohesiveness in MSR,
i.e., MSR values are widely discrepant among different genres and are inherently cohesive within a same genre.
The ambiguity of MSR is also diversified for different musical genres.
Besides, the MSR values of a given musical genre are symmetric in terms of the range boundaries (i.e., mean) as well as the ambiguity of range boundaries (i.e., standard deviation).
The regression lines from original tempos of music pieces to MSR values ($\alpha_{min}$ and $\alpha_{max}$) are almost horizontal within a given musical genre in the experiments,
which further indicates that MSR has little correlation with tempo unlike that with musical genre.
Besides, we also analyze the MSR-based similarity between musical genres by computing the overlapping area of covering regions on the $\alpha_{min}$--$\alpha_{max}$ coordinate system.
This new measuremeant of similarity offers a new perspective on musical genres based on human perception of music stretching resistance.
MSR is a psychoacoustic reflection of human perception of music~\cite{speedmelody:sysmu08},
and the study on MSR also sheds new light on content-aware music adaption~\cite{musicresize:mms13,musicresize:mm12,musicresize:icme10}
and dynamic attending theory~\cite{regularity:mathbehavior02,dynamicattending:psychreview89,rhythmicattending:cognition00,dynamicattending:psychreview99}.

\section*{Methods}

\subsection*{Participants}

We recruited 17 college students as participants in the experiments.
These participants ranged in age from 18 to 25 years old, with $29.4\%$ female and $70.6\%$ male.
Most of them were non-musicians except for two participants who had received piano education for more than 3 years by the time of experiments.
It has been proved that listeners (musicians or non-musicians) can make consistent judgments on whether music pieces are played overly fast or overly slow~\cite{tempopercept:perception06,memorymanipulation:icmpc06}.
The composition of participants in our experiments is similar to the real situation in our daily lives.
The participants were selected among the people who enjoyed listening to music and were willing to spend at least half an hour every day to conduct the experiments.
The experiments last for about one month so that the experimental results would be less influenced by the short-term changes of participants' physical or mental states,
such as moods (happy, sad), time (morning, evening), locations (home, workplace), and weather (rainy, sunny).
Since the participants were expected to conduct the listening experiments for a long period of time,
we excluded those short-term participants who could not make through the one-month experiments to get the results from the final 17 participants.
Although the final number of participants is not very large, we tried to minimize the impact of personal preferences by increasing the overlap of music pieces that different participants listened,
and majority-voting the results they reported.
All the participants were paid a little bit for their endeavor when the experiments finished.

\subsection*{Experimental Settings}

\begin{figure*}[t]
\centering
\includegraphics[width=0.6\textwidth]{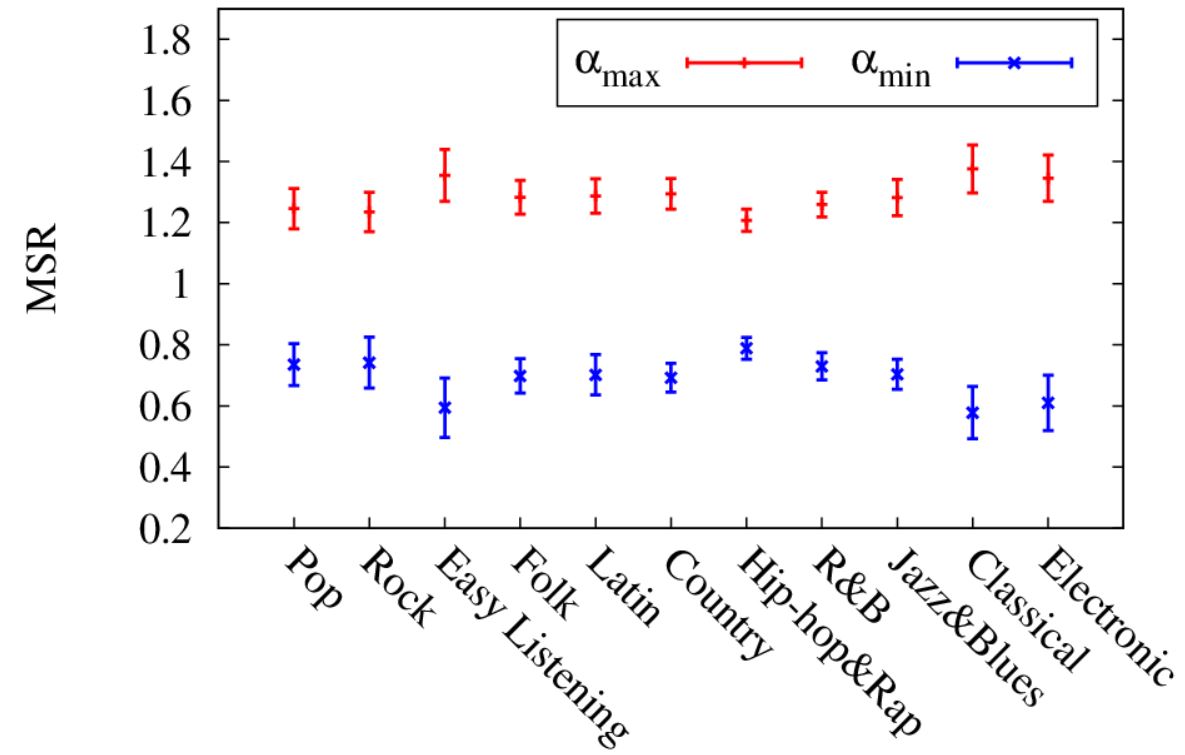}
\caption{
The error bars of MSR distribution of 11 different musical genres.}
\label{fig_genre_tree_values}
\end{figure*}

We used the collection of 894 songs from~\cite{msr:spl13} in this study.
These songs identically cover 11 musical genres as shown in Table~\ref{tb_msrvalues_slopes}.
These songs were randomly crawled from a music website~\footnote{http://www.top100.cn},
and the genres of these songs were annotated by referring to the meta data of these songs as well as the genre taxonomy from Wikipedia and some popular music websites.

All the songs in the collection were stretched in time domain using the synchronized overlap-add (SOLA) method~\cite{sola:icassp85} to avoid pitch shift,
which was implemented in the SoundTouch library~\footnote{http://www.surina.net/soundtouch}.
Each song was stretched into different versions of discrete stretching rates between $(0.00,2.00)$ with a rate step 0.02.
That is, each song has 49 compressed versions with stretching rates in $\{0.98,0.96,\-...,0.02\}$ as well as 49 elongated versions with stretching rates in $\{1.02,1.04,...,\-1.98\}$.
Please note that no music piece with stretching rate beyond 2.00 will be acceptable from our experience
since the elongation would have destroyed the musical structure too much and made the elongated music piece sound uncomfortable.

Although there are some other different music stretching methods~\cite{musicresize:mms13,musicresize:mm12,musicresize:icme10,musicrescale:eurographics13}, we choose SOLA~\cite{sola:icassp85} in this work because SOLA is a more fundamental work in the music stretching literature and it uniformly stretches music pieces.
Therefore, the results reported in this paper are about uniformly music stretching.

\subsection*{Procedure}

Each participant was delivered at least one package of 20 random songs from our collection as well as the 98 stretched versions (49 compressed + 49 elongated) for each song.
The genre composition in each song package is random,
and some participants were delivered more song packages since they were faster in conducting the listening experiments.
The participants were asked to listen to these stretched versions one after another,
and judge whether each stretched version is acceptable.
They could choose the order of stretching rates to listen in their preferred ways as long as they could give judgement on $\alpha_{min}$ and $\alpha_{max}$.
We offered the participants a few general judging criteria,
e.g., speed acceptance, lyrics density acceptance and overall listening acceptance.
The degradation of sound quality after stretch using SOLA method also turns out to
be an important factor to influence people's psychoacoustic acceptance.
For example, over elongation will make the sound interruptive
while over compression will mess up the audio signals and make it sound noisy.
Besides the given judging criteria, it is up to the participants to make their own judgement based on the comfort they feel in the listening.
Though the total number of stretched versions of each song was not small,
the participant did not have to listen to all of them,
and they only needed to locate the minimum and the maximum acceptable stretching rates of each song in their package(s).
To further help the participants to judge the acceptance of the stretched versions,
we also developed a new music player for them, which supported `one-click' switch between two stretched versions of a same song at the same position, e.g., $30\%$ of the song~\cite{resic:mmm14}. %
With this music player, it would be easier for the participants to judge whether the listening version is over-stretched at the currently playing segment compared with the original version or the other acceptable versions.

When the participants completed the tasks in the delivered package(s),
(s)he input the results, i.e., the $\alpha_{min}$ and the $\alpha_{max}$ of each song in the package, into our experimental web pages.
Specifically, the participants enter the values of $\alpha_{min}$ (e.g. 0.82) and $\alpha_{max}$ (e.g. 1.26) of the songs in their package(s) through text fields on the web page.
The participants were allowed to correct their results before the final submission by the end of the one-month experiment.
Thus, the minimum compressing rate $\alpha_{min}$ and the maximum elongating rate $\alpha_{max}$ of each song in our collection were obtained for later analysis.

\section*{Results and Discussion}

We performed the analysis of variance (ANOVA) on the MSR values obtained from the listening experiments.
MSR values represent $\alpha_{min}$ and $\alpha_{max}$ of each song reported by the participants.
The results show that MSR values are significantly affected by musical genres
(F-test, alpha levels are the $11$ genres, $\alpha_{min}$: $F(10, 883)=74.683$, $P<0.001$,
$\alpha_{max}$: $F(10, 883)=56.407$, $P<0.001$).
Next, the relationship between MSR and musical genres is discussed in detail based on the experimental results.

\subsection*{Basic MSR Properties}

\begin{table*}[t]
\centering
\begin{tabular}{c|c|c|c|r|r}
    \hline
    Genre       &  \#.Piece   &   $\alpha_{min}$   &    $\alpha_{max}$     &   Slope $\alpha_{min}$   &  Slope $\alpha_{max}$     \\
    \hline
    Pop         &  $87$	      &   $0.735\pm0.069$  &    $1.246\pm0.066$    &   $2.78\times10^{-4}$	  &  $-5.30\times10^{-5}$       \\
    Rock 	    &  $95$       &   $0.742\pm0.083$  &    $1.235\pm0.064$    &   $6.85\times10^{-6}$    &	 $-1.45\times10^{-4}$     \\
    Easy Listening	& $83$	  &   $0.594\pm0.097$  &    $1.355\pm0.086$    &   $-3.15\times10^{-4}$	  &  $6.75\times10^{-5}$      \\
    Folk	    &  $80$	      &   $0.698\pm0.057$  &    $1.283\pm0.056$    &   $-4.29\times10^{-5}$   &  $-1.47\times10^{-4}$     \\
    Latin	    &  $68$ 	  &   $0.702\pm0.066$  &    $1.287\pm0.056$    &   $-1.34\times10^{-4}$	  &  $-1.25\times10^{-5}$     \\
    Country	    &  $86$	      &   $0.692\pm0.047$  &    $1.294\pm0.050$    &   $1.73\times10^{-6}$ 	  &  $1.79\times10^{-4}$      \\
    Hip-hop\&Rap & $82$	      &   $0.789\pm0.036$  &    $1.207\pm0.036$    &   $6.98\times10^{-5}$	  &  $3.12\times10^{-6}$      \\
    R\&B	    &  $90$	      &   $0.729\pm0.045$  &    $1.259\pm0.040$    &   $-2.77\times10^{-4}$	  &  $-1.71\times10^{-5}$     \\
    Jazz\&Blues	&  $78$	      &   $0.703\pm0.049$  &    $1.282\pm0.059$    &   $-4.11\times10^{-5}$   &	 $-1.24\times10^{-4}$     \\
    Classical	&  $73$	      &   $0.578\pm0.085$  &    $1.376\pm0.079$    &   $-4.50\times10^{-4}$	  &  $4.89\times10^{-4}$      \\
    Electronic	&  $72$	      &   $0.610\pm0.090$  &    $1.346\pm0.076$    &   $-7.79\times10^{-5}$	  &  $5.31\times10^{-5}$      \\
    \hline
\end{tabular}
\caption{The means and deviations of the MSR values of different musical genres,
as well as the slopes of regression lines from the original tempo of a music piece to $\alpha_{min}$/$\alpha_{max}$ within each musical genre.}
\label{tb_msrvalues_slopes}
\end{table*}

Fig.~\ref{fig_genre_tree_values} illustrates the error bars (means and standard deviations) of MSR values for the 11 musical genres (the statistics are shown in Table~\ref{tb_msrvalues_slopes}),
from which we can draw the following four basic conclusions about MSR:
\begin{itemize}
\item \textbf{Inter-Genre Discrepancy}: Wide discrepancies in MSR values under different genres are observed.
        The position and the stretch of MSR of musical genres are quite different.
        For instance, \emph{Easy Listening}, \emph{Electronic} and \emph{Classical} music have a wider stretching range (the interval between the mean of $\alpha_{min}$ and that of $\alpha_{max}$),
        while \emph{Hip-hop}\&\emph{Rap} and \emph{R}\&\emph{B} exhibit a narrower stretching reach.
        It is in line with the character effect (fast and slow) of music pieces on human preferred speeds and ranges of acceptability~\cite{speedmelody:sysmu08}.
\item \textbf{Intra-Genre Cohesiveness}: MSR values are substantiated to be cohesive under a given genre according to the significance of musical genres in the analysis of variance.
\item \textbf{Ambiguity}: The ambiguity of MSR is greater for \emph{Easy Listening}, \emph{Electronic} and \emph{Classical} music pieces seen from the standard deviations.
    The larger the standard deviation is, the larger ambiguity the MSR of a musical genre is.
    This probably results from the rhythmic features of these musical genres since there may hardly be any fixed rhythmic patterns for these aforementioned genres,
    for instance, a piece of piano music or violin music.
    On the contrary, songs like \emph{R}\&\emph{B}, \emph{Hip-hop}\&\emph{Rap} usually follow a solid tempo throughout the whole music piece.
\item \textbf{Symmetry}: Under a given genre, MSR tends to be symmetric between $\alpha_{min}$ and $\alpha_{max}$,
    on both the range boundaries (mean) and the ambiguity of range boundaries (standard deviation).
    This property comes from the symmetric criteria that listeners used to judge $\alpha_{min}$ and $\alpha_{max}$.
    Supposedly, the optimal tempo of a given song should occur near its base one.
    As a result, if a song is over-compressed and sounds uncomfortable,
    it is more likely that the elongated one with the same shift of stretching rate increase will also sound uneasy, and vice versa.
\end{itemize}

\subsection*{Intra-Genre Linear Regression With Tempo}

The stretching operations on a given music piece will lead to an inversely proportional relationship, $t_s=\frac{t_o}{r}$,
where $r$ is the stretching rate between the range ($0.00$, $2.00$),
$t_o$ and $t_s$ are the tempos of the original music piece and that of the stretched version measured by beat per minute, respectively.
Evidently, $t_o$ is fixed for a given music piece.
Thus, $t_s$ meets the upper limit when $r$ equals $\alpha_{min}$,
while $t_s$ reaches the floor boundary when $r$ equals $\alpha_{max}$.
So as to identify the relationship between boundary tempos and MSR,
the linear regression is performed from base tempos to $\alpha_{min}$ and $\alpha_{max}$ in the music collection, respectively.
This is to study whether or not songs with different tempos under a given musical genre would generally have different $\alpha_{min}$ and $\alpha_{max}$.
The regression lines are almost horizontal under all genres since their slopes are very close to zero (Table~\ref{tb_msrvalues_slopes}).
For example, the steepest slope in Table~\ref{tb_msrvalues_slopes} is $-4.89 \times 10^{-4}$ of $\alpha_{max}$ of \emph{Classical Music}.
Since the original tempo of a music piece mostly varies between $[0,200]$ BPM, which can only cause less than 0.1 bias in the $\alpha_{max}$ from other music pieces of this genre.
Consequently, for a given musical genre, music pieces with different base tempos usually have similar MSR,
which verifies the intra-genre cohesiveness of MSR from another point of view.
Under a given genre, the limits of $r$ are fixed,
and thus the upper and the floor boundaries of $t_s$ vary according to the value of $t_o$ of a given music piece.
The fact that the regressed slopes within each musical genre are almost zero is also a solid proof that MSR has little correlation with tempo unlike that with musical genre.

\subsection*{MSR-based Musical Genre Similarity}

\begin{figure*}[t]
\centering
\includegraphics[width=0.9\textwidth]{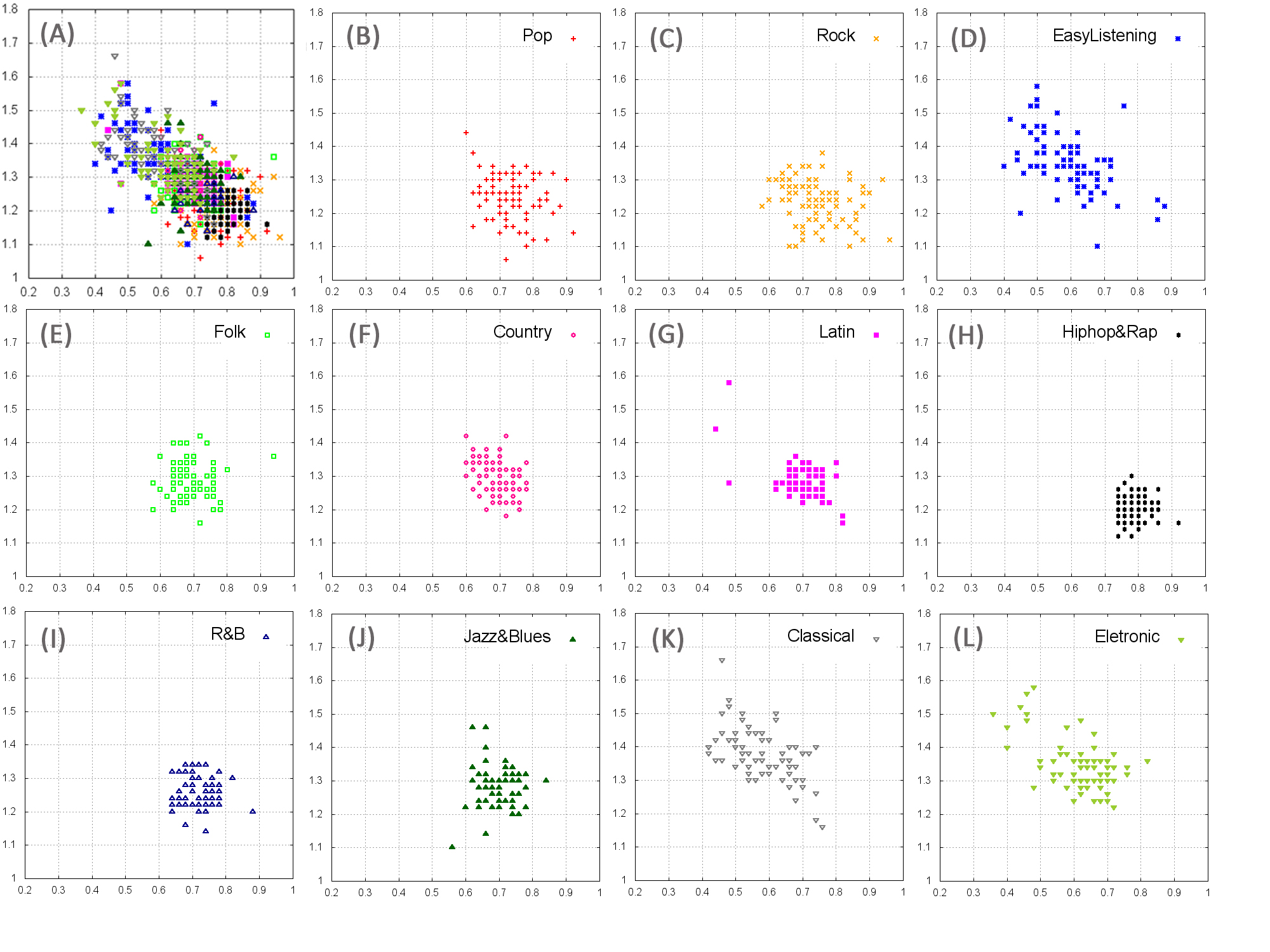}
\caption{The scatter plots of the MSR value distribution within each musical genre.
The abscissa and the ordinate of each point in the panels represent the values of $\alpha_{min}$ and $\alpha_{max}$, respectively.
(A) is the assembly of the points from (B) to (L). }\label{fig_scatterplot_msr}
\end{figure*}

\begin{figure*}[h]
\centering
\includegraphics[width=0.9\textwidth]{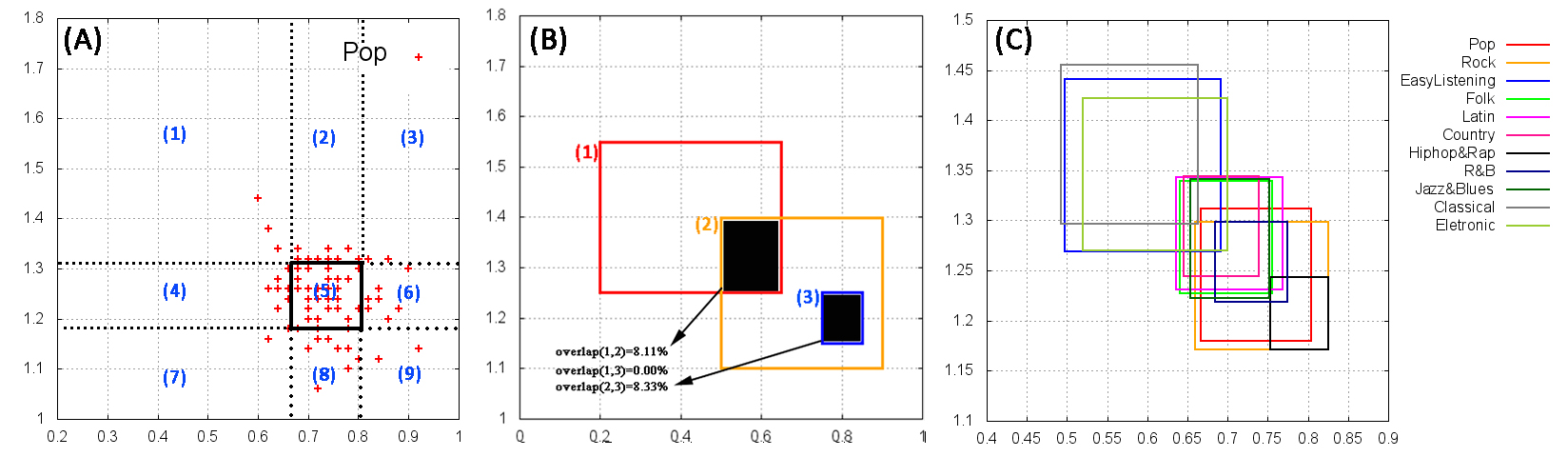}
\caption{The MSR-based similarity of musical genres.
(A) The MSR rectangle of \emph{Pop} music containing points on the $\alpha_{min}-\alpha_{max}$ coordinate system.
(B) An example of the possible relationship between MSR rectangles.
(C) The illustration of the MSR rectangles of all musical genres.}
\label{fig_msr_area_overlap}
\end{figure*}

To further investigate the relationship between MSR and musical genres,
we illustrate the scatter plots of the MSR values within each musical genre in Fig.~\ref{fig_scatterplot_msr}.
Each point in these panels represents a pair $(\alpha_{min},\alpha_{max})$,
while each point may correspond to a few songs sharing the same MSR values.
Obviously, the covered region of the points within each musical genre diversifies.
For example, the points in Fig.~\ref{fig_scatterplot_msr}d cover a wide range which means the ambiguity/variance of MSR in this group is high,
however, the points in Fig.~\ref{fig_scatterplot_msr}h are very close with each other spreading in a small range,
and thus the ambiguity/variance of MSR in this group is relatively lower in contrast.
The assembly of these points in Fig.~\ref{fig_scatterplot_msr}a also shows the difference in the MSR value distribution of different musical genres,
which inspires us that we can distinguish different musical genres based on their covering area of the points on the $\alpha_{min}-\alpha_{max}$ coordinate system.

We use a rectangle to mark the edges of the MSR region (a.k.a.~MSR rectangle) on the $\alpha_{min}$-$\alpha_{max}$ coordinate system for a given genre.
The rectangle is located by the following four coordinates:
\begin{equation}
(Mean(\alpha_{min}) \pm StdDev (\alpha_{min}), Mean(\alpha_{max}) \pm StdDev(\alpha_{max})),
\end{equation}
where $Mean(\alpha_{min})$ and $StdDev(\alpha_{min})$ represent the mean and the standard deviation of $\alpha_{min}$ in the given musical genre, respectively.
So are $Mean(\alpha_{max})$ and $StdDev(\alpha_{max})$.
These values are shown in Table~\ref{tb_msrvalues_slopes}.
Fig.~\ref{fig_msr_area_overlap}a shows the MSR rectangle of \emph{Pop} music.
The larger the MSR rectangle is, the more ambiguity of MSR the given musical genre has.
The MSR rectangle divides the $\alpha_{min}-\alpha_{max}$ coordinate system into 9 parts on the first quadrant as illustrated in Fig.~\ref{fig_msr_area_overlap}a.
Part 9 can be considered as the `safe' stretching area of the given musical genre
since the points in this region are within the acceptable stretching range.
Parts 1,2,3,4,7 are the `dangerous' stretching areas of the given musical genre by contrast.
Furthermore, Parts 5,6,8 can therefore be considered as the transition areas from the `dangerous' stretching region to the `safe' stretching region.
If the music stretching tasks~\cite{musicresize:mms13,musicresize:mm12,musicresize:icme10,wsola:icassp93,ola:sc00} are performed within the `safe' region,
the stretched results will be the most likely to be accepted by general audience.
On the contrary, if stretched in the `dangerous' region, the results will usually be unacceptable for general audience.
Since different musical genres lead to different MSR rectangles,
these MSR rectangles can be used to distinguish different musical genres from the perspective of MSR
which has never been studied in the literature.

\begin{table}
\centering
\scriptsize
\begin{tabular}{c|ccccccccccc}
    \hline
    Genre          &  Pop           & Rock          & Easy Listening & Folk          &  Latin        & Country       & Hip-hop\&Rap   &   R\&B        &  Jazz\&Blues  &  Classical    &  Electronic \\
    \hline
    Pop            &  \textbf{1.0}  & 0.713         & 0.021          & 0.323         & 0.334         & 0.219         & 0.159          & \textbf{0.395}& 0.346         &  \emph{0.0}   & 0.032       \\
    Rock           &                & \textbf{1.0}  & 0.018          & 0.255         & 0.259         & 0.168         & \textbf{0.244} & \textbf{0.339}& 0.275         &  1.7e-4       & 0.025       \\
    Easy Listening &                &               & \textbf{1.0}   & 0.082         & 0.092         & 0.088         & \emph{0.0}     & 0.005         & 0.063         &  0.658        & 0.748       \\
    Folk           &                &               &                & \textbf{1.0}  & 0.808         & 0.675         & 0.002          & 0.344         & 0.822         &  0.024        & 0.113       \\
    Latin          &                &               &                &               & \textbf{1.0}  & 0.625         & 0.009          & 0.351         & 0.692         & 0.031         & 0.125       \\
    Country        &                &               &                &               &               & \textbf{1.0}  & \emph{0.0}     & 0.223         & 0.648         & 0.024         & 0.125       \\
    Hip-hop\&Rap   &                &               &                &               &               &               & \textbf{1.0}   & 0.042         & \emph{0.0}    & \emph{0.0}    & \emph{0.0}  \\
    R\&B           &                &               &                &               &               &               &                & \textbf{1.0}  & 0.380         & \emph{0.0}    & 0.014       \\
    Jazz\&Blues    &                &               &                &               &               &               &                &               & \textbf{1.0}  & 0.010         & 0.092       \\
    Classical      &                &               &                &               &               &               &                &               &               & \textbf{1.0}  & 0.492       \\
    Electronic     &                &               &                &               &               &               &                &               &               &               & \textbf{1.0} \\
    \hline
\end{tabular}
\normalsize
\caption{The MSR-based similarity between musical genres.
The similarity matrix is symmetric, and only half the matrix is shown.
The inclusion relationship is in \textbf{bold} mode and the exclusion relationship is in \emph{italic} mode.
The others are in the intersection relationship.}
\label{tb_msrbased_similarities}
\end{table}

The area of MSR rectangle, i.e., $4 \times\-StdDev(\alpha_{min})\-\times StdDev(\alpha_{max})$ can be used to measure the relative MSR ambiguity of musical genres.
Moreover, the relationship of MSR rectangles from different musical genres falls into three categories:
(1) \textbf{Inclusion} --- one rectangle is totally included in another rectangle,
or two rectangles are exactly the same;
(2) \textbf{Exclusion} --- two rectangles are totally disjoint;
(3) \textbf{Intersection} --- two rectangles have the common overlap as well as the disjoint part.
Fig.~\ref{fig_msr_area_overlap}b shows an example of the possible relationship between musical genres based on their MSR rectangles.
\emph{Rect} 1 is intersected with \emph{Rect} 2, while \emph{Rect} 3 is included in \emph{Rect} 2 and excluded from \emph{Rect} 1.
The overlaps of rectangles are filled with black,
and we can compute the overlap ratio using the Jaccard similarity:
\begin{equation}\label{lb_msr_similarity}
Sim(Rect_i,Rect_j)=\frac{Rect_i \wedge Rect_j}{Rect_i \vee Rect_j}.
\end{equation}
$Sim(Rect_i,Rect_j)$ measures how much similarity the musical genres that $Rect_i$ and $Rect_j$ represent, share from the perspective of MSR.
The relationships of MSR rectangles of all musical genres are shown in Fig.~\ref{fig_msr_area_overlap}c.
We can see that all the inclusion, the exclusion and the intersection relationship exist among the musical genres presented in this paper.

Next, we computed the MSR-based similarity between musical genres using Eq.~\ref{lb_msr_similarity}, and the results are shown in Table~\ref{tb_msrbased_similarities}.
Since the similarity matrix is symmetric, only half the matrix is shown.
Obviously, the similarity between two musical genres whose MSR rectangles satisfy the \emph{exclusion} relationship is zero (in \emph{italic} mode).
The \emph{inclusion} relationship (in \textbf{bold} mode) is found that
\emph{R}\&\emph{B}$\subseteq$\emph{Pop},
\emph{R}\&\emph{B}$\subseteq$\emph{Rock},
and \emph{Hip-hop}\&\emph{Rap}$\subseteq$\emph{Rock}
where $\subseteq$ is the inclusion operator.
Most of the relationship observed is \emph{intersection}.
However, the ratio of overlap in the intersection relationship varies between different pairs of musical genres.
The MSR-based similarity offers a new look to explore the relationship between musical genres.
We can see from Table~\ref{tb_msrbased_similarities} that high MSR-based similarity is observed between some musical genre pairs, e.g., \emph{Pop}-\emph{Rock}, \emph{Folk}-\emph{Latin}, \emph{Folk-Jazz}\&\emph{Blues}.
This kind of similarity may be very difficult or even impossible to study using meta-data, audio content or cultural backgrounds
since MSR-based similarity is related to human psychoacoustic perception of music as well as people's degrees of self-adaption to changes of music such as tempo, event density and lyrics density.

We believe that our work has made prospective attempts in studying the relationship between MSR and musical genres,
and more issues remain to be further investigated in the future.
Our findings on the MSR hold not only on digital music recordings which are stretched using complex signal processing algorithms~\cite{musicresize:mms13,musicresize:mm12,musicresize:icme10,wsola:icassp93,ola:sc00}.
It can also be applied in live music performance, for example,
the acceleration and the deceleration of live piano or violin performances to accompany singers in a concert.


\begin{thebibliography}{10}
\expandafter\ifx\csname url\endcsname\relax
  \def\url#1{\texttt{#1}}\fi
\expandafter\ifx\csname urlprefix\endcsname\relax\def\urlprefix{URL }\fi
\providecommand{\bibinfo}[2]{#2}
\providecommand{\eprint}[2][]{\url{#2}}

\bibitem{wsola:icassp93}
\bibinfo{author}{Verhelst, W.} \& \bibinfo{author}{Roelands, M.}
\newblock \bibinfo{title}{An overlap-add technique based on waveform similarity
  (wsola) for high quality time-scale modification of speech}.
\newblock In \emph{\bibinfo{booktitle}{IEEE international conference on
  acoustics, speech and signal processing}}, \bibinfo{pages}{554--557}
  (\bibinfo{year}{1993}).

\bibitem{ola:sc00}
\bibinfo{author}{Verhelst, W.}
\newblock \bibinfo{title}{Overlap-add methods for time-scaling of speech}.
\newblock \emph{\bibinfo{journal}{Speech Communication}}
  \textbf{\bibinfo{volume}{30}}, \bibinfo{pages}{207--221}
  (\bibinfo{year}{2000}).

\bibitem{msr:spl13}
\bibinfo{author}{Chen, J.} \& \bibinfo{author}{Wang, C.}
\newblock \bibinfo{title}{Automatic music stretching resistance classification
  using audio features and genres}.
\newblock \emph{\bibinfo{journal}{IEEE Signal Processing Letters}}
  \textbf{\bibinfo{volume}{20}}, \bibinfo{pages}{1249--1252}
  (\bibinfo{year}{2013}).

\bibitem{musicresize:mms13}
\bibinfo{author}{Liu, Z.}, \bibinfo{author}{Wang, C.}, \bibinfo{author}{Wang,
  J.}, \bibinfo{author}{Wang, H.} \& \bibinfo{author}{Bai, Y.}
\newblock \bibinfo{title}{Adaptive music resizing with stretching, cropping and
  insertion}.
\newblock \emph{\bibinfo{journal}{Multimedia System}}
  \textbf{\bibinfo{volume}{19}}, \bibinfo{pages}{359--380}
  (\bibinfo{year}{2013}).

\bibitem{musicresize:mm12}
\bibinfo{author}{Liu, Z.}, \bibinfo{author}{Wang, C.}, \bibinfo{author}{Bai,
  Y.}, \bibinfo{author}{Wang, H.} \& \bibinfo{author}{Wang, J.}
\newblock \bibinfo{title}{Musiz: a generic framework for music resizing with
  stretching and cropping}.
\newblock In \emph{\bibinfo{booktitle}{ACM Multimedia}},
  \bibinfo{pages}{523--532} (\bibinfo{year}{2011}).

\bibitem{musicresize:icme10}
\bibinfo{author}{Liu, Z.}, \bibinfo{author}{Wang, C.}, \bibinfo{author}{Guo,
  L.}, \bibinfo{author}{Bai, Y.} \& \bibinfo{author}{Wang, J.}
\newblock \bibinfo{title}{Lydar: a lyrics density based approach to
  non-homogeneous music resizing}.
\newblock In \emph{\bibinfo{booktitle}{IEEE international conference on
  multimedia and expo}}, \bibinfo{pages}{310--315} (\bibinfo{year}{2010}).

\bibitem{musicrescale:eurographics13}
\bibinfo{author}{Wenner, S.}, \bibinfo{author}{Bazin, J.-C.},
  \bibinfo{author}{Sorkine-Hornung, A.}, \bibinfo{author}{Kim, C.} \&
  \bibinfo{author}{Gross, M.}
\newblock \bibinfo{title}{Scalable music: Automatic music retargeting and
  synthesis}.
\newblock \emph{\bibinfo{journal}{Eurographics}} \textbf{\bibinfo{volume}{32}},
  \bibinfo{pages}{345--354} (\bibinfo{year}{2013}).

\bibitem{memorymanipulation:icmpc06}
\bibinfo{author}{Brennan, D.} \& \bibinfo{author}{Stevens, C.}
\newblock \bibinfo{title}{The effect of pitch, tempo and proportional pitch and
  tempo manipulation on memory of familiar musical excerpts}.
\newblock In \emph{\bibinfo{booktitle}{International conference on music
  perception and cognition}}, \bibinfo{pages}{1771--1778}
  (\bibinfo{year}{2006}).

\bibitem{voiceselective:nature00}
\bibinfo{author}{Berlin, P.}, \bibinfo{author}{Zatorre, R.~J.},
  \bibinfo{author}{Lafaille, P.}, \bibinfo{author}{Ahad, P.} \&
  \bibinfo{author}{Pike, B.}
\newblock \bibinfo{title}{Voice-selective areas in human auditory cortex}.
\newblock \emph{\bibinfo{journal}{Nature}} \textbf{\bibinfo{volume}{403}},
  \bibinfo{pages}{309--312} (\bibinfo{year}{2000}).

\bibitem{speedinmusic:jasa10}
\bibinfo{author}{Madison, G.} \& \bibinfo{author}{Paulin, J.}
\newblock \bibinfo{title}{Ratings of speed in real music as a function of both
  original and manipulated beat tempo}.
\newblock \emph{\bibinfo{journal}{Journal of the Acoustical Society America}}
  \textbf{\bibinfo{volume}{128}}, \bibinfo{pages}{3032--3040}
  (\bibinfo{year}{2010}).

\bibitem{speedmelody:sysmu08}
\bibinfo{author}{Bisesi, E.} \& \bibinfo{author}{Vicario, G.~B.}
\newblock \bibinfo{title}{Psychoacoustic aspects of the speed of melody
  performance}.
\newblock In \emph{\bibinfo{booktitle}{International conference of students of
  systematic musicology}}, \bibinfo{pages}{7--11} (\bibinfo{year}{2008}).

\bibitem{regularity:mathbehavior02}
\bibinfo{author}{Large, E.} \& \bibinfo{author}{Palmer, C.}
\newblock \bibinfo{title}{Perceiving temporal regularity in music}.
\newblock \emph{\bibinfo{journal}{Mathematical Behavior}}
  \textbf{\bibinfo{volume}{26}}, \bibinfo{pages}{1--37} (\bibinfo{year}{2002}).

\bibitem{dynamicattending:psychreview89}
\bibinfo{author}{Jones, M.} \& \bibinfo{author}{Boltz, M.}
\newblock \bibinfo{title}{Dynamic attending and responses to time}.
\newblock \emph{\bibinfo{journal}{Psychological Review}}
  \textbf{\bibinfo{volume}{96}}, \bibinfo{pages}{459--491}
  (\bibinfo{year}{1989}).

\bibitem{rhythmicattending:cognition00}
\bibinfo{author}{Drake, C.}, \bibinfo{author}{Jones, M.} \&
  \bibinfo{author}{Baruch, C.}
\newblock \bibinfo{title}{The development of rhythmic attending in auditory
  sequences: attunement, referent period, focal attending}.
\newblock \emph{\bibinfo{journal}{Cognition}} \textbf{\bibinfo{volume}{77}},
  \bibinfo{pages}{251--288} (\bibinfo{year}{2000}).

\bibitem{dynamicattending:psychreview99}
\bibinfo{author}{Large, E.} \& \bibinfo{author}{Jones, M.}
\newblock \bibinfo{title}{The dynamics of attending: how people track
  time-varying events}.
\newblock \emph{\bibinfo{journal}{Psychological Review}}
  \textbf{\bibinfo{volume}{106}}, \bibinfo{pages}{119--159}
  (\bibinfo{year}{1999}).

\bibitem{tempopercept:icmpc04}
\bibinfo{author}{Moelants, D.} \& \bibinfo{author}{Mckinney, M.~F.}
\newblock \bibinfo{title}{Tempo perception and musical content: What makes a
  piece fast, slow or temporally ambiguous?}
\newblock In \emph{\bibinfo{booktitle}{International conference on music
  perception and cognition}}, \bibinfo{pages}{558--562} (\bibinfo{year}{2004}).

\bibitem{resonancetheory:cim04}
\bibinfo{author}{Mckinney, M.~F.} \& \bibinfo{author}{Moelants, D.}
\newblock \bibinfo{title}{Deviations from the resonance theory of tempo
  induction}.
\newblock In \emph{\bibinfo{booktitle}{International conference on
  interdisciplinary musicology}}, \bibinfo{pages}{124--125}
  (\bibinfo{year}{2004}).

\bibitem{msrbound:nime04}
\bibinfo{author}{Lee, E.}, \bibinfo{author}{Nakra, T.~M.} \&
  \bibinfo{author}{Borchers, J.}
\newblock \bibinfo{title}{You're the conductor: a realistic interactive
  conducting system for children}.
\newblock In \emph{\bibinfo{booktitle}{International conference on new
  interfaces for musical expression}}, \bibinfo{pages}{68--73}
  (\bibinfo{year}{2004}).

\bibitem{tempopercept:perception06}
\bibinfo{author}{Quinn, S.} \& \bibinfo{author}{Watt, R.}
\newblock \bibinfo{title}{The perception of tempo in music}.
\newblock \emph{\bibinfo{journal}{Perception}} \textbf{\bibinfo{volume}{35}},
  \bibinfo{pages}{267--280} (\bibinfo{year}{2006}).

\bibitem{genreclassify:spm06}
\bibinfo{author}{Scaringella, N.}, \bibinfo{author}{Zoia, G.} \&
  \bibinfo{author}{Mlynek, D.}
\newblock \bibinfo{title}{Automatic genre classification of music content: a
  survey}.
\newblock \emph{\bibinfo{journal}{IEEE Signal Processing Magazine}}
  \textbf{\bibinfo{volume}{23}}, \bibinfo{pages}{133--141}
  (\bibinfo{year}{2006}).

\bibitem{genreclassify:sigir03}
\bibinfo{author}{Li, T.}, \bibinfo{author}{Ogihara, M.} \& \bibinfo{author}{Li,
  Q.}
\newblock \bibinfo{title}{A comparative study on content-based music genre
  classification}.
\newblock In \emph{\bibinfo{booktitle}{SIGIR}}, \bibinfo{pages}{282--289}
  (\bibinfo{year}{2003}).

\bibitem{genreclassify:tsp02}
\bibinfo{author}{Tzanetakis, G.}, \bibinfo{author}{Essl, G.} \&
  \bibinfo{author}{Cook, P.}
\newblock \bibinfo{title}{Automatic musical genre classification of audio
  signals}.
\newblock \emph{\bibinfo{journal}{IEEE Transaction on Speech and Audio
  Processing}} \textbf{\bibinfo{volume}{10}}, \bibinfo{pages}{293--302}
  (\bibinfo{year}{2002}).

\bibitem{genreclassify:spl07}
\bibinfo{author}{Bagci, U.} \& \bibinfo{author}{Erzin, E.}
\newblock \bibinfo{title}{Automatic classification of musical genres using
  inter-genre similarity}.
\newblock \emph{\bibinfo{journal}{IEEE Signal Processing Letters}}
  \textbf{\bibinfo{volume}{14}}, \bibinfo{pages}{521--524}
  (\bibinfo{year}{2007}).

\bibitem{sola:icassp85}
\bibinfo{author}{Roucos, S.} \& \bibinfo{author}{Wilgus, A.}
\newblock \bibinfo{title}{High quality time-scale modification for speech}.
\newblock In \emph{\bibinfo{booktitle}{IEEE international conference on
  acoustics, speech and signal processing}}, \bibinfo{pages}{493--496}
  (\bibinfo{year}{1985}).

\bibitem{resic:mmm14}
\bibinfo{author}{Chen, J.} \& \bibinfo{author}{Wang, C.}
\newblock \bibinfo{title}{RESIC: A tool for music stretching resistance
  estimation}.
\newblock In \emph{\bibinfo{booktitle}{MultiMedia Modeling}}, Lecture Note in
  Computer Science, \bibinfo{pages}{386--389} (\bibinfo{year}{2014}).

\end{thebibliography}

\section*{Acknowledgements}

We would like to thank all the volunteers who participated in the listening experiments for their contributions which form the basis of this paper.

\end{document}